\documentclass[prl,twocolumn,preprintnumbers,reprint]{revtex4-1} 
\usepackage{dcolumn,bm}
\usepackage{amsmath,amssymb,amsfonts,graphics,graphicx,amscd,amsfonts,color}

\begin{document}

\title{Deconfinement transition as black hole formation by the condensation of QCD strings}

\author{Masanori Hanada$^{a,b,c}$}      
\author{Jonathan Maltz$^{d}$} 
\author{Leonard Susskind$^{a}$}
\affiliation{
${}^a$ Stanford Institute for Theoretical Physics, Stanford University, Stanford CA 94305, USA}
\affiliation{
${}^b$
Yukawa Institute for Theoretical Physics, Kyoto University, Kitashirakawa Oiwakecho, Sakyo-ku, Kyoto 606-8502, Japan 
}
\affiliation{
${}^c$
The Hakubi Center for Advanced Research, Kyoto University, Yoshida-Ushinomiya-cho, Sakyo-ku, Kyoto 606-8501, Japan
  }
  \affiliation{
${}^d$  Kavli Institute for the Physics and Mathematics of the Universe, Todai Institutes for Advanced Study, The University of Tokyo, Kashiwa, Chiba 277-8582, Japan
}

\preprint{YITP-14-35}
\preprint{SU-ITP-14/10}
\preprint{IPMU14-0106}

\begin{abstract}
We argue that the deconfinement transition of large-$N$ Yang-Mills theory is 
the condensation of very long and self-intersecting chromo-electric flux strings (QCD string), 
which is analogous to the formation of a black hole in string theory. We do this by using lattice gauge theory and matrix models. 
As evidence, we derive an analytic formula for the deconfinement temperature in the strong coupling limit and confirm it numerically. 
Dual gravity descriptions interpreted in this manner should make it possible to understand 
the details of the formation of black holes in terms of fundamental strings. 
We argue that very simple matrix models capture the essence of the formation of black holes.
\end{abstract}

\maketitle

{\it Introduction.---}
In the gauge/gravity duality \cite{Maldacena:1997re}, 
the deconfinement transition of the gauge theory is dual to the formation of a black hole in the gravity side \cite{Witten:1998zw}. 
In this paper, we give an intuitive way of understanding this correspondence, without referring to a sophisticated dictionary of the duality. 
Our argument does not assume the dual gravity description, and hence it is applicable to generic gauge theories including QCD.  
We pay attention to the stringy degrees of freedom in gauge theory -- the Wilson loops -- and see how their behaviors change across the deconfinement transition. 
The authors believe that this result will be useful in understanding the formation and thermalization of the quark-gluon plasma in QCD, and the formation of a black hole in the gravitational picture, from a unified perspective. 

\begin{figure}[htb]
\begin{center} 
\scalebox{0.5}{
\includegraphics[width=9cm]{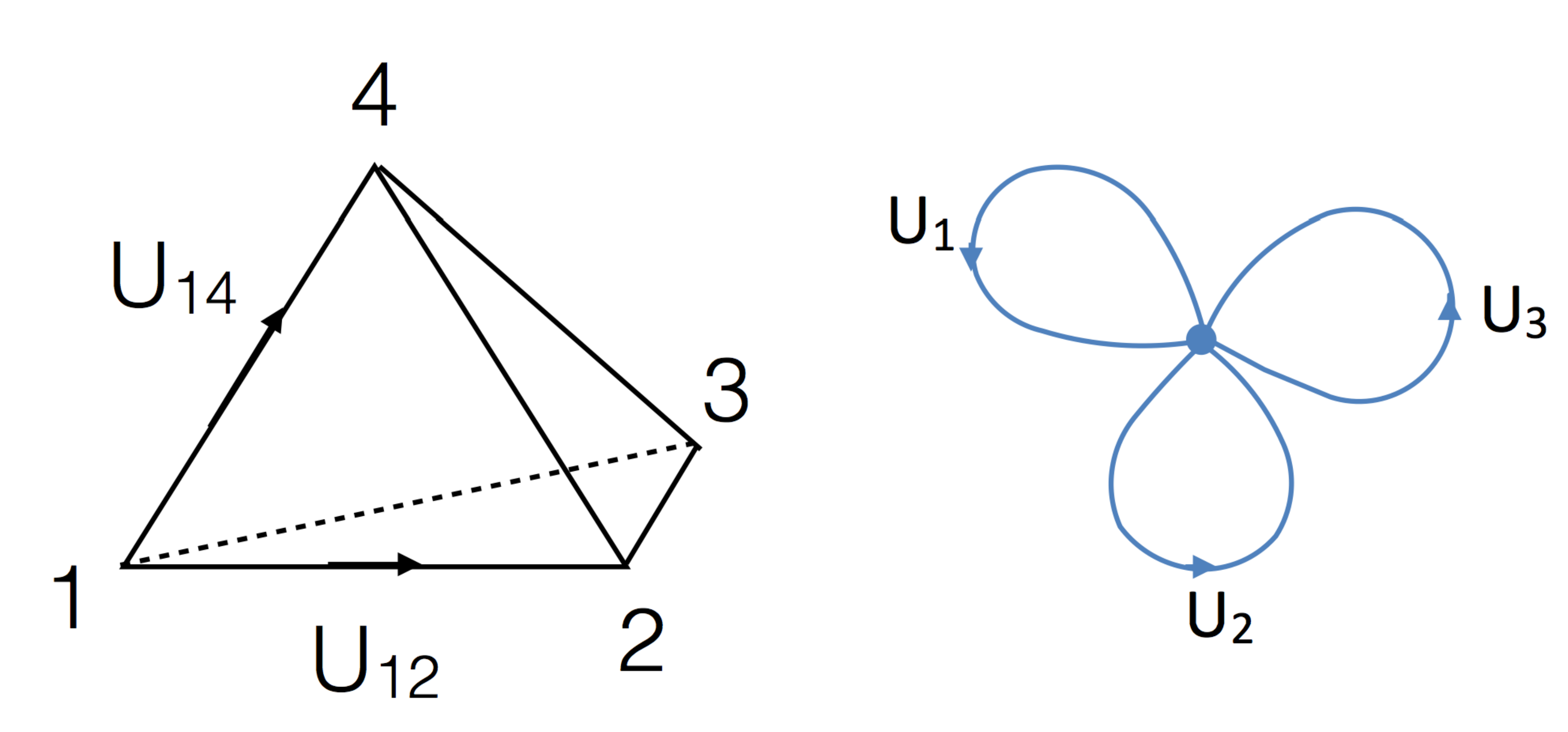}}
\end{center}
\caption{
Tetrahedron matrix model (left) and single-site model (Eguchi-Kawai model) with three links (right).
}
\label{fig:tetrahedron}
\end{figure}

Although our argument applies to rather generic gauge theories and matrix models, our initial motivation comes from 
a simple matrix model for the black hole \cite{Susskind:2013aaa,Susskind:2014rva} to study the duality \cite{Witten:1998zw} 
between the deconfinement transition in 4d ${\cal N}=4$ super Yang-Mills theory on a three-sphere and the Hawking-Page transition \cite{Hawking:1982dh} 
of a black hole in the AdS space. 
At the Hawking-Page transition, a small black hole whose Schwarzschild radius is of order AdS radius is formed. 
Such a small black hole can naturally be identified with a long and winding string \cite{Susskind:1993ws,Horowitz:1996nw}.  
One of the authors (L.~S.) has argued that it can be modeled by using a lattice gauge theory with a continuum time and a few spatial lattice sites, 
e.g. a tetrahedron (Fig.~\ref{fig:tetrahedron}) sitting at the center of the AdS space, by identifying the string and chromo-electric flux string. 
Our result in this paper justifies this argument. 

As a concrete example, let us consider pure $U(N)$ Yang-Mills theory. 
The Hilbert space is spanned by Wilson loops acting on the vacuum $|0\rangle$, 
$W_C W_{C'}\cdots |0\rangle$,  
where $W_C$ represent the Wilson loop along a closed contour $C$. 
In the standard identification of the Feynman diagrams and the string world-sheet \cite{'tHooft:1973jz}, 
the Wilson loop is naturally interpreted as the creation operator of the string. 

In the large-$N$ limit and at sufficiently strong coupling, the energy of the string is approximated by its length. 
In the confinement phase, the energy is of order $N^0$ per unit volume, 
and hence a typical state is a finite-density gas of loops with finite length. 
In this gas, two loops can intersect with each other and combined to form a longer string. 
Alternatively, when a loop intersects with itself, it can be split into two shorter loops. 
However such joining and splitting are suppressed at large-$N$. 

In the deconfinement phase, the energy density is of order $N^2$. In this phase the loops necessarily intersect 
$O(N^2)$ times. Although the interaction at each intersection is $1/N$-suppressed, 
small interactions at many intersections accumulate to a non-negligible amount. 
As we will explain shortly,  a typical state consists of finitely many very long and self-intersecting strings, 
whose lengths are of $O(N^2)$. 
In the string theory, it is natural to interpret such very long and self-intersecting strings as a black hole \cite{Susskind:1993ws,Horowitz:1996nw}.  
In this sense, {\it the deconfinement transition can be understood as the formation of a `black hole' through 
condensation of QCD strings} \cite{comment}.  

\begin{figure}[htb]
\begin{center} 
\scalebox{0.35}{
\includegraphics[width=12cm]{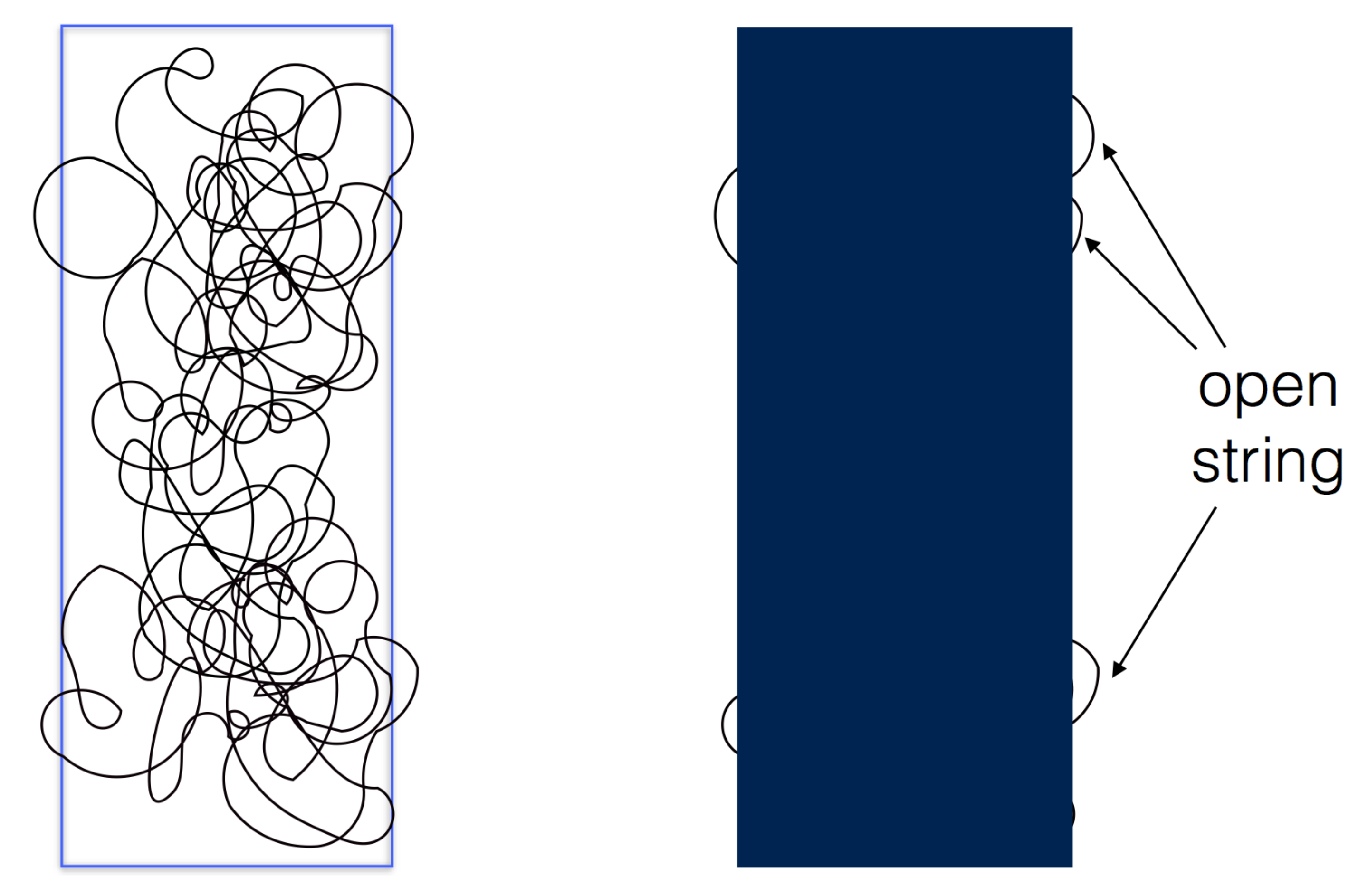}}
\caption{
Closed string picture (left) and open string picture (right).  
}\label{fig:BlackBrane}
\end{center}
\end{figure}

When we identify the long string with a black hole, fluctuations of the string near the horizon are regarded as open strings 
attached to black hole (Fig.~\ref{fig:BlackBrane}). In terms of the gauge theory, these open strings are open Wilson lines  
which have $N$ color degrees of freedom at their endpoints. Therefore, we can interpret that the black hole is made from $N$ D-branes. 

\begin{figure}[htb]
\begin{center} 
\scalebox{0.5}{
\includegraphics[width=12cm]{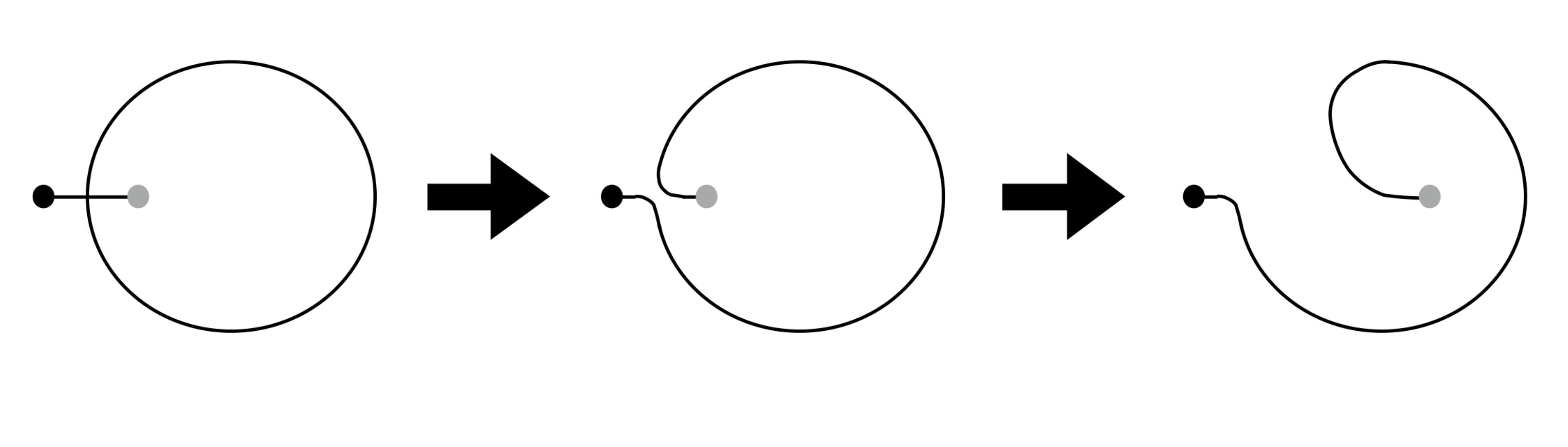}}
\caption{
Deconfinement of a pair of probe quark and anti-quark. 
}\label{fig:probe_deconfinement}
\end{center}
\end{figure}

In the Euclidean theory, the deconfinement is characterized by the condensation of the Polyakov loop, 
which can naturally be related to the black hole geometry in the gravity dual \cite{Witten:1998zw,Aharony:2003sx}. 
This can be explained from our picture as follows. First notice that the condensation of the Polyakov loop is equivalent to 
the disappearance of the linear confinement potential between a pair of probe quark and anti-quark. 
In terms of strings, the linear potential in the confining phase appears because an open string connecting probes 
must be stretched as they are separated. In the deconfining phase, however, as soon as  a short open string is introduced, 
it intersects with closed strings many times, and they immediately interact at one of the intersections to form a long open string 
(Fig.~\ref{fig:probe_deconfinement}). Therefore, probes can be separated without stretching the open string, or equivalently, without costing energy. 
Note that the interaction is crucial for this argument; the deconfining phase cannot be described by free strings \`{a} la Hagedorn.     

In the following, we demonstrate this picture by using lattice gauge theory and matrix models as concrete examples. 
When it is applied to theories with gravity duals, the `analogy' turns to an equivalence. That is to say that the  
`QCD string' is naturally identified with the fundamental string, and their condensation is equivalent to the formation of an actual black hole 
in superstring theory.


{\it Quantitative argument on lattice.--- }
Let us consider $(D+1)$-dimensional $U(N)$ Yang-Mills lattice gauge theory in the Hamiltonian formulation \cite{Kogut:1974ag}. 
(We consider $D\ge 2$, which goes through a first-order deconfinement transition as the temperature is raised.)  
The time direction is continuous, and $D$-dimensional space is regularized by a $D$-dimensional square lattice. 
The Hamiltonian is given by 
 the kinetic term (or electric term) $K$ and the potential term (magnetic term) $V$ respectively, 
$H=K + V$, 
where 
\begin{eqnarray}\label{kineticone}
K = \frac{\lambda N}{2}\sum_{\vec{x}}\sum_{\mu}\sum_{\alpha=1}^{N^2}
\left(E^\alpha_{\mu,\vec{x}}\right)^2
\nonumber
\end{eqnarray}
and  
\begin{eqnarray}
V = \frac{N}{\lambda}\sum_{\vec{x}}
\sum_{\mu<\nu}
\left(
N-
{\rm Tr} (U_{\mu,\vec{x}}U_{\nu,\vec{x}+\hat{\mu}}U^\dagger_{\mu,\vec{x}+\hat{\nu}}U^\dagger_{\nu,\vec{x}})
\right).
\nonumber
\end{eqnarray}
Here $U_\mu(\vec{x})$ is the unitary link variable connecting $\vec{x}$ and $\vec{x}+\hat{\mu}$, 
where $\hat{\mu}$ is the unit vector along $\mu$-direction. 
 $E^{\alpha}_{\mu,\vec{x}}$ is defined via its commutation relation which is given by,
\begin{eqnarray}
& &
[E^\alpha_{\mu,\vec{x}},U_{\nu,\vec{y}}] = \delta_{\mu\nu}\delta_{\vec{x}\vec{y}}\cdot \tau^\alpha U_{\nu,\vec{y}},  
\nonumber\\
& &
[E_{\mu,\vec{x}},E_{\nu,\vec{y}}] 
=
[U_{\mu,\vec{x}},U_{\nu,\vec{y}}] 
=
[U_{\mu,\vec{x}},U^\dagger_{\nu,\vec{y}}] 
=
0. 
\nonumber
\end{eqnarray}
$\tau_\alpha$ ($\alpha=1,2,\cdots,N^2$) are generators of the $U(N)$ group, which are $N\times N$ Hermitian matrices
normalized as 
$\sum_{\alpha=1}^{N^2}\tau^\alpha_{ij} \tau^\alpha_{kl} = \delta_{il}\delta_{jk}/N^2$. 
The vacuum $|0\rangle$ is annihilated by $E^\alpha_{\mu,\vec{x}}$, 
$E^\alpha_{\mu,\vec{x}}|0\rangle=0$.

The Hilbert space consists of gauge-invariant states, in which Wilson loops acting on $|0\rangle$ form an over-complete basis.  
In the lattice gauge theory, the Wilson loop is the trace of the product of link variables along a contour $C$,   
$W_C=Tr\left(U_{\mu,\vec{x}}U_{\nu,\vec{x}+\hat{\mu}}\cdots U_{\rho,\vec{x}-\hat{\rho}}\right)$. 
The Wilson loop sources a set of interacting gluons. The correspondence between the Feynman diagram expansion 
and the string world-sheet leads us to interpret the Wilson loop as the creation operator of the closed string.  
 
Let us say the Wilson loop is not self-intersecting when each link variable appears only once. 
Similarly, two loops $W_C$ and $W_{C'}$ do not intersect when they do not share the same link variable. 
(With this definition, even if two loops go through the same point $\vec{x}$, they do not necessarily `intersect'. 
This is actually a natural generalization of the intersection between strings 
in continuum space, because two strings represented by the Wilson loops on the lattice can interact only when they share the same link, as we will see shortly.) 

When the loop $W_{C}$ does not self-intersect, the electric term $K$ acts on the state $|W_C\rangle\equiv W_C|0\rangle$ as 
\begin{eqnarray}
K|W_C\rangle
=
\frac{\lambda N}{2}\sum_{\mu,\vec{x},\alpha}
\left[
E^\alpha_{\mu,\vec{x}},
\left[
E^\alpha_{\mu,\vec{x}},W_C
\right]
\right]|0\rangle
=
\frac{\lambda L}{2}|W_C\rangle,   
\nonumber
\end{eqnarray}
where $L$ is the length of the contour $C$ in the lattice unit. 
In the same manner, for a multi-loop state $|W_C, W_{C'}, \cdots\rangle = W_C W_{C'}\cdots|0\rangle$, 
$K$ acts as $K|W_C, W_{C'}, \cdots\rangle =\frac{\lambda}{2}(L+L'+\cdots)|W_C, W_{C'}, \cdots\rangle$ 
when there is no intersection. 

When two loops $W_C$ and $W_{C'}$ intersect once by sharing $U_{\mu,\vec{x}}$, they can be joined to form a longer loop as follows. 
Let us write $W_C$ and $W_{C'}$ as $W_C={\rm Tr}(U_{\mu,\vec{x}}M)$ and $W_{C'}={\rm Tr} (U_{\mu,\vec{x}}M')$, 
where $M$ and $M'$ are product of other link variables. 
Then, 
\begin{eqnarray}\label{twollong}
\lefteqn{K|W_C, W_{C'}\rangle }
\nonumber\\
&=&
\frac{\lambda(L+L')}{2}|W_C, W_{C'}\rangle
\nonumber\\
& &
+
\lambda N
\sum_\alpha {\rm Tr}(\tau^\alpha U_{\mu,\vec{x}}M)\cdot {\rm Tr}(\tau^\alpha U_{\mu,\vec{x}}M')|0\rangle
\nonumber\\
&=&
\frac{\lambda(L+L')}{2}|W_C, W_{C'}\rangle
+
\frac{\lambda}{N}
 {\rm Tr}(U_{\mu,\vec{x}}M U_{\mu,\vec{x}}M')|0\rangle.  
 \nonumber
\end{eqnarray}
The second term is a longer string whose length is $L+L'$. 
In the same manner, a self-intersecting string can be split into two strings. 
If there are multiple intersections, such joinings and splittings take place at all intersections.

In this manner, the electric term $K$ contains all the essence for the black hole formation -- 
it measures the total length of the strings, and also joins and splits strings.  
The magnetic term $V$ is the plaquette, which is the smallest possible Wilson loop; this term adds a very short string (one plaquette) to the states. 
This term is not negligible in the continuum limit and quantitative argument can change. In particular, the string tension is not necessarily proportional 
to the coupling constant at UV. 

\begin{figure}[htb]
\begin{center} 
\scalebox{0.4}{
\includegraphics[width=12cm]{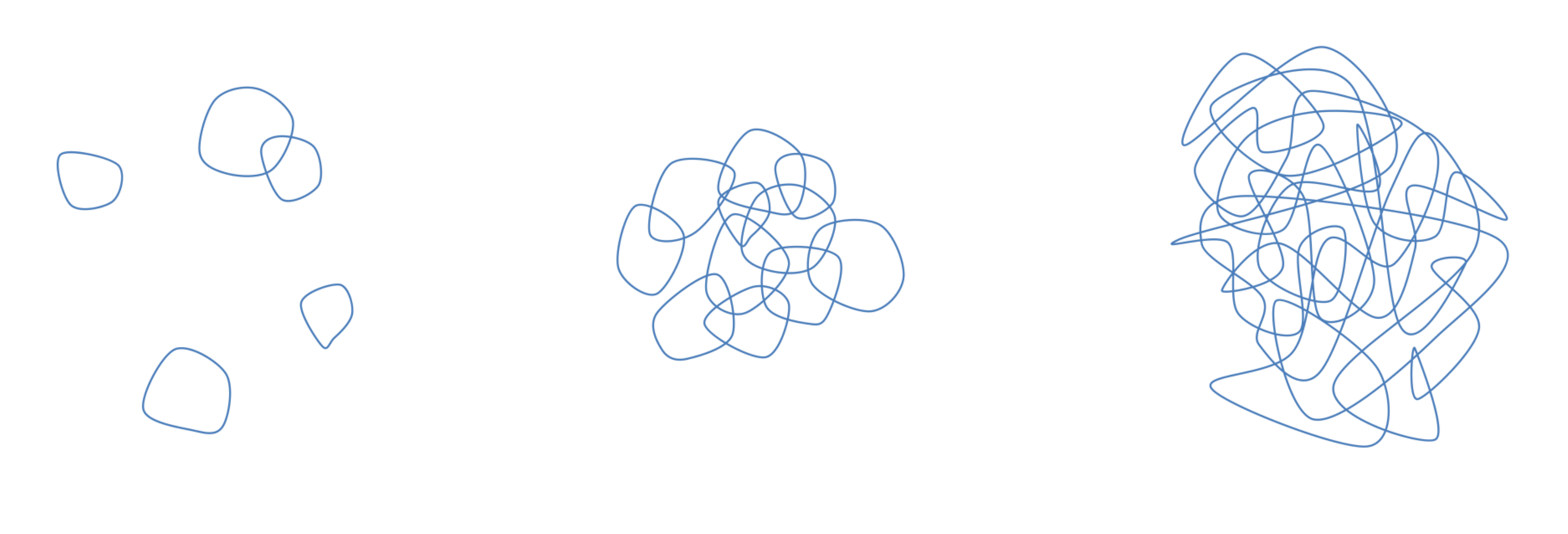}}
\caption{
The confining phase is the sparse gas of free strings (left). 
As the system is heated up, the gas become denser (middle),  
the interaction becomes non-negligible and a long string is formed (right).  
}\label{fig:deconfinement_transition}
\end{center}
\end{figure}

{\it Confinement phase as gas of free strings.---}
In the confining phase, the energy density is of order $N^0$. Therefore, the length of the strings is also $O(N^0)$. 
Then joining and splitting coming from the electric term is negligible at large-$N$, because there are only $O(N^0)$ intersections,  
and interaction at each intersection is $1/N$-suppressed. 
This phase can therefore be understood as a gas of non-interacting strings. 
(The left of Fig.~\ref{fig:deconfinement_transition})

{\it Deconfinement phase as black hole.--}
In the deconfinement phase, the energy density and hence the total length of the strings are of order $N^2$.  
The strings must intersect $O(N^2)$ times, and then the $1/N$-suppressed interaction at each intersection accumulates and becomes non-negligible 
(Fig.~\ref{fig:deconfinement_transition}). 
Strings are joining, splitting, as well as changing their shapes very rapidly in this phase. 

There are many string states with total length $L_{total} \sim N^2$. 
One extreme, are cases with configurations containing a single long string, while the other extreme, are states only containing the $O(N^2)$ shortest strings (plaquettes). 
A single long string can possess a lot of states; after $U_{\mu,\vec{x}}$, one can multiply any link variables originating from 
the point $\vec{x}+\hat{\mu}$ except $U_{\mu,\vec{x}}^\dagger$, and hence at each point there are $2D-1$ choices, 
and the entropy is roughly $\log( (2D-1)^{L_{total}}) = L_{total}\log(2D-1)\sim N^2$.  
The states which consist of several long strings can carry the same amount of entropy. 
On the other hand, if we consider a bunch of very short strings, say a bunch of plaquetts, 
then the entropy is just of order $\log N$, 
because the state can be specified by the number of each type of plaquette. 
Therefore, it is natural to conclude that typical states contain a few long strings. 
In string theory, this is exactly how the black hole (or more precisely a $D$-dimensional black brane) is formed from the fundamental string. 

In order to justify this picture, we derive the following analytic predictions, and then confirm them numerically.
Let us consider the strong coupling limit ($\lambda=\infty$, $V=0$). 
Then the energy of the deconfinement phase is proportional to 
the total length of the strings $L_{total}(T)$, which is an unknown function of the temperature $T$, 
because the electric term measures the length,   
$E=K=\frac{\lambda}{2}L_{total}(T)$. 
The entropy $S$ is also proportional to $L_{total}(T)$,  
$S=L_{total}\log(2D-1)$.   
Therefore the free energy $F=E-TS$ is given by 
$F=L_{total}(T)  ((\lambda/2)-T\log(2D-1))$. 
The confinement phase is favored when it is positive and 
the deconfinement transition takes place when it crosses zero, 
because the free energy of the confinement phase is zero up to a $1/N$ correction. 
Therefore, although we do not know the $T$-dependence of the length $L_{total}$, 
we can easily determine the critical temperature $T_c$, 
\begin{eqnarray}
T_c=\lambda/(2\log(2D-1)). 
\label{eq:Tc_strong_coupling}
\end{eqnarray} 
Note that this derivation is formally the same as the derivation of the Hagedorn temperature in free string theory on a lattice (see e.g. \cite{Patel:1983sc}). 
 

{\it Matrix models.---}
Strictly speaking, the $(D+1)$-dimensional Yang-Mills theory is analogous to a $D$-dimensional black brane rather than a black hole, 
because the condensation of the strings fills entire $D$-dimensional space. 
In order to describe a black hole ($0$-brane), let us consider the dimensionally reduced $D$-matrix model 
$H=K + V$, 
where 
$K = \frac{\lambda N}{2}\sum_{\mu}\sum_{\alpha=1}^{N^2}
\left(E^\alpha_\mu\right)^2$ 
is the kinetic term (or electric term) and  
$V = \frac{N}{\lambda}
\sum_{\mu\neq\nu}
\left(
N-
{\rm Tr} (U_{\mu}U_{\nu}U^\dagger_{\mu}U^\dagger_{\nu})
\right)$.
Now the space is reduced to a single point, and $D$ link variables are attached to that point 
(see Fig.~\ref{fig:tetrahedron} 
for the case of $D=3$).  
This is the Eguchi-Kawai model \cite{Eguchi:1982nm} with a continuous time direction. 
At strong coupling the $U(1)^D$ center symmetry along the spatial directions, 
$U_\mu\to e^{i\theta_\mu}U_\mu$, is not spontaneously broken.  
Then this theory is known to be equivalent to the $(D+1)$-dimensional theory at large-$N$  
in the sense 
that translationally invariant observables in the latter, for example the energy density and entropy density, 
are reproduced from the former to the leading order in the $1/N$-expansion \cite{Eguchi:1982nm}. 
At weak coupling, this model is analogous to the bosonic part of the matrix model of M-theory \cite{Banks:1996vh}, 
which is dual to the black zero-brane in type IIA supergravity in the 't Hooft large-$N$ limit \cite{Itzhaki:1998dd}. 
For $D\ge 2$, this theory exhibits the deconfinement transition. 
At sufficiently strong coupling, the transition is of first order.  
In the deconfinement phase, the energy and the entropy is of order $N^2$, 
and typical state should be described by long, winding strings such as $Tr(U_1U_2U_1^\dagger U_1^\dagger U_2U_2U_1\cdots)$. 
All of these arguments parallel the case of the $(D+1)$-dimensional lattice, and 
\eqref{eq:Tc_strong_coupling} should hold in this case as well.  

We can also consider other matrix models such as the tetrahedron model (Fig.~\ref{fig:tetrahedron}), 
which consists of four points. 
There are six link variables $U_{mn}$ ($m\neq n$) 
which satisfy $U_{mn}^\dagger=U_{nm}$. 
We have shown numerically that this model also possesses a first-order deconfinement transition. 
The long string can be described with a term such as $Tr(U_{12}U_{23}U_{31}U_{14}U_{42}\cdots)$. 
The entropy is $S=L_{total}\log 2$, and \eqref{eq:Tc_strong_coupling} is modified as 
$T_c=\lambda/(2\log 2)$.


{\it Numerical confirmation.---}
Thanks to the Eguchi-Kawai equivalence \cite{Eguchi:1982nm}, we do not have to study the $(D+1)$-dimensional YM theory, 
and hence we consider the matrix models. 
In order to study the thermodynamics, we consider the theory in Euclidean time, and compactify the time direction 
to a circle with circumference $\beta=1/T$. The Lagrangian in Euclidean signature is given by 
$L=K+V$, 
where $V$ is the same as before and 
$K = \frac{N}{2\lambda}\sum_{\vec{x}}\ {\rm Tr}
\left(
D_tU_{\vec{x}}\cdot (D_tU_{\vec{x}})^\dagger
\right)$ 
with $D_t U_{\vec{x}}=\partial_t U_{\vec{x}} - i[A_t,U_{\vec{x}}]$. 
We regularize the model by introducing $n_t$ lattice sites along the temporal direction. 
The action is 
\begin{eqnarray}
S_{lattice}
&=&
-\frac{N}{2a\lambda}\sum_{\mu,t} 
{\rm Tr}\left(V_tU_{\mu,t+a}V_{\mu,t}^\dagger U_{\mu,t}+c.c.
\right)
\nonumber\\
& &
+
\frac{aN}{\lambda}
\sum_{\mu\neq\nu,t}
\left(
N-
{\rm Tr} (U_{\mu,t}U_{\nu,t}U^\dagger_{\mu,t}U^\dagger_{\nu,t})
\right)  
\nonumber
\end{eqnarray}
where $t=a,2a,\cdots,n_ta$, where $a=\beta/n_t$ is the lattice spacing, and $t=n_ta$ is identified to $t=0$. 
$V_t$ is the unitary link variable connecting $t$ and $t+a$, and $U_{\mu,t}$ are spatial links at time $t$. 
The tetrahedron matrix model can be regularized in a similar manner.

\begin{figure}[htb]
\begin{center}
\scalebox{0.7}{
\rotatebox{0}{ 
\includegraphics[width=12cm]{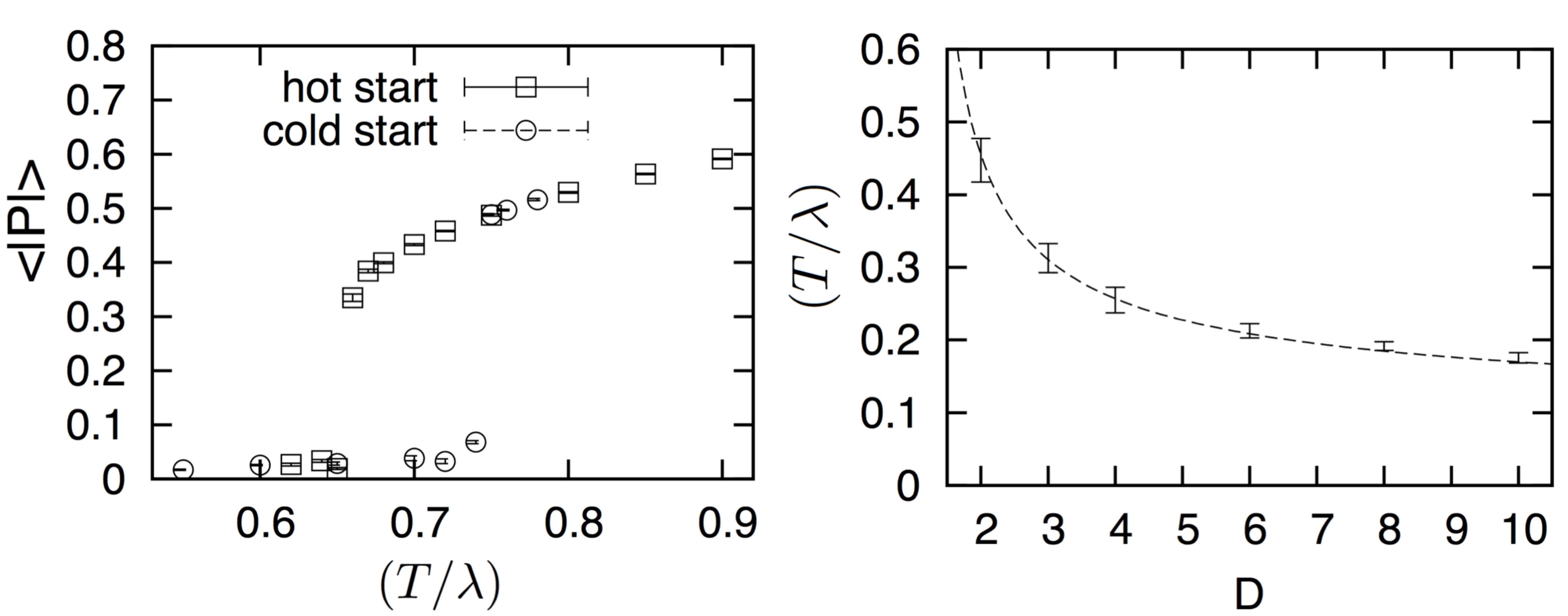} }}
\end{center}
\caption{
(Left)The expectation value of the Polyakov loop in the tetrahedron matrix model in the strong coupling limit. 
$N=64, n_t=12$. 
(Right)The temperature range of the hystereses of the Eguchi-Kawai model.   
The dashed line is the analytic prediction for the critical temperature, $(T_c/\lambda)=1/(2\log(2D-1))$. 
$N=64, n_t=16$.  
}
\label{fig:plot_deconfinement}
\end{figure}

In the following we concentrate on the strong coupling limit, where the magnetic terms are omitted. 
The order parameter for the deconfinement transition is the Polyakov loop, 
which is defined by $P=\frac{1}{N}{\rm Tr}(V_{t=a}V_{t=2a}\cdots V_{t=n_ta})$ for the Eguchi-Kawai model.  
Because the expectation value of $P$ itself vanishes trivially due to the $U(1)$ phase, we consider $\langle|P|\rangle$. 
We performed hot and cold start simulations, in which the temperature is gradually decreased and increased, respectively.  
In Fig.~\ref{fig:plot_deconfinement}, $\langle|P|\rangle$ in the tetrahedron model is plotted. 
We can see a clear hysteresis at $0.66\le (T/\lambda)\le 0.74$, which means the transition is of first order as expected. 
The theoretically predicted critical temperature $(T_c/\lambda)=1/(2\log 2)\simeq 0.721$ is in this range.

The plot show that the confining phase characterized by the vanishing Polakov loop survives as a metastable state above $T_c$, 
which is the same as the Hagedorn temperature for the free strings. 
Such a behavior is possible because the free-string picture is not valid when the string becomes longer;  
the potential barrier separating two phases arises due to a very nontrivial interaction, 
and it can survive beyond $T_c$. Still it would be possible to see the Hagedorn behavior as a $1/N$ correction.

We observed similar hystereses also in the Eguchi-Kawai model with various values of $D$. 
In the right panel of Fig.~\ref{fig:plot_deconfinement} we plot the temperature range of the hystereses for $D=2,3,\cdots,10$. 
We can see that the theoretical predictions for $T_c$ is consistent with the simulation results.  

{\it Discussion.---}
In the gauge theories with dual gravity descriptions, 
the condensation of the Wilson loops argued in this paper is equivalent to the condensation of fundamental strings 
and formation of an actual black hole. 
By following the time evolution of the loops, it should be possible to see how a black hole forms and thermalizes. 
As we have seen, qualitative features are common even in the strong coupling limit of simple matrix models, 
and hence they should serve as good toy models for a black hole. 

Among various possible applications of this work is the fast scrambling conjecture \cite{Sekino:2008he},   
which claims that a black hole and large-$N$ gauge theories are the fastest 
scramblers of information.
The intuitive picture discussed in this paper should be useful for understanding the microscopic mechanism of 
fast scrambling from gauge theory. 
Clearly, a huge number of simultaneous interactions at various intersections should be the essence of fast scrambling. 
The understanding of fast scrambling in gauge theory is interesting from the point of view of the quantum information theory, 
and even more, would be useful in understanding the very fast thermalization of RHIC fireball \cite{Heinz:2004pj}. 
We hope to report progress in this direction in near future. 

{\it Acknowledgements.---}
The authors would like to thank D.~Anninos, E.~Berkowitz, S.~Hartnoll, P.~Hayden, S.~Hellerman, A.~Hook, 
K.~Schmitz, S.~Shenker, H.~Shimada, C.~Siegel, D.~Stanford, T.~Takayanagi  and N.~Yamamoto 
for stimulating discussions and comments. 
M.~H. was supported in part by the Ministry of Education, Science, 
Sports and Culture, Grant-in-Aid for Young Scientists (B), 
25800163. 
J.~M. was supported by World Premier International Research Center Initiative (WPI Initiative), MEXT, Japan. 
Support for the research of L.~S. came through NSF grant Phy-1316699 and the Stanford Institute for Theoretical Physics.


\bibliography{bibliography}

\end{document}